\documentclass[%
 reprint,
showpacs,
 amsmath,amssymb,
 aps,
prb,
]{revtex4-1}

\usepackage{graphicx}
\usepackage{dcolumn}
\usepackage{bm}

\begin{document}

\preprint{APS/123-QED}

\title{Effect of Dipolar Interaction in Molecular Crystals}

\author{Danh-Tai Hoang}%
 \email{danh-tai.hoang@u-cergy.fr}
\author{H. T. Diep}
\email{diep@u-cergy.fr, corresponding author}
 \affiliation{%
Laboratoire de Physique Th\'eorique et Mod\'elisation,
Universit\'e de Cergy-Pontoise, CNRS, UMR 8089\\
2, Avenue Adolphe Chauvin, 95302 Cergy-Pontoise Cedex, France.\\
 }%




\date{\today}

\begin{abstract}
We  investigate in this paper the ground state and the nature of the  transition from an orientational ordered phase at low temperature  to the disordered state at high temperature in a molecular crystal.   Our model is a Potts model which takes into account the exchange interaction $J$ between nearest-neighbor molecules and  a dipolar interaction between molecular axes in three dimensions.    The dipolar interaction is characterized by  two parameters: its amplitude $D$ and the cutoff distance $r_c$.    If the molecular axis at a lattice site has  three orientations, say the $x$, $y$ or $z$ axes, then when $D=0$,  the system is equivalent to the 3-state Potts model: the transition to the disordered phase is known to be of first order. When $D\neq 0$, the ground-state configuration is shown to be composed of two independent interpenetrating layered subsystems which form a sandwich whose periodicity depends on $D$ and $r_c$.  We show  by extensive Monte Carlo simulation with a histogram method that the phase transition remains of first order at relatively large values of $r_c$.
\begin{description}
\item[PACS numbers: 61.30.Cz, 64.70.M-]
\end{description}
\end{abstract}

\pacs{Valid PACS appear here}
\maketitle


\section{Introduction}
The nature of the transition from one phase to another is one of the most important problems in statistical physics and in various areas of materials science not limited to physics.  Since the introduction of the renormalization group \cite{Wilson,Zinn} with new concepts based on the system symmetry and scale, the understanding of the nature of the phase transition in many systems  becomes clear.  However, there are complicated systems where that method encounters much of difficulties in application.  One can mention frustrated systems where competing interactions cause highly degenerate ground states with instabilities \cite{Diep2005}.

In this paper, we are interested in the phase transition in molecular crystals which has been a subject of intensive investigations since 40 years.  This spectacular development was due to numerous
 applications of liquid crystals in daily life \cite{deGennes,Chandrasekhar}.  Liquid crystals are somewhere between solid and liquid states where molecules have some spatial orientations which, under some conditions, can order themselves into some structures such as nematic and smectic phases.  Liquid crystal display uses properties of these ordered phases \cite{Fukuda,Takezoe}.   In spite of the large number of applications using experimental findings, theoretical understanding in many points is still desirable.   One of the most important aspects of the problem is the origin of layered structures observed in smectic ordering: in smectic phases the molecules are ordered layer by layer with periodicity.  Experiments have discovered, for example, smectic phases of 3-layer, 4-layer, or 6-layer periodicity \cite{Mach1,Mach2,Johnson,Hirst,Wang}. Different theories have been suggested to interpret these observations \cite{Miyashita,Cepic,Olson,Dolganov2003,Emelyanenko,Fernandes,Dolganov2008,Dolganov2010}.
One of the unanswered questions is what is the origin of the long-period layered structure in observed smectic phases?
Hamaneh et al. \cite{Hamaneh} suggested that the  effective long-range interaction
 is due to  bend fluctuations of the smectic layers which may stabilize commensurate structures in particular the six-layer phase.  This suggestion is too qualitative to allow a clear understanding of the long-period layered structure in smectic phases.  To our opinion, we should distinguish long-range correlation and long-range interaction. We know that long-range correlation can be a consequence of short-range interactions, not necessarily of  long-range ones, as it is well known in theory of  critical phenomena \cite{Zinn}. However, a long-range correlation does not imply a particular long-period  configuration.

 To look for a physical origin of long-period structures, we concentrate ourselves in the present paper to the effect of  a dipolar interaction in a Potts model, in addition to an exchange interaction between nearest neighbors (NN).   We suppose that each molecule has a molecular axis which can lie on one of the three principal directions.  An example of such a molecule is the ammoniac molecule NH$_3$.  Without the dipolar interaction, the interaction between neighboring molecules gives rise to an orientational order of molecular axes at low temperature.  In this situation, the system can be described by a 3-state Potts model in three dimensions (3D).
The  Potts model is very important in statistical physics.  It is in fact a class of models each of which is defined by the number of  states $q$ that an individual particle or molecule can have.  The interaction between two neighboring molecules is negative if they are in the same state, it is zero otherwise.   Exact solutions are found for many Potts models in  two dimensions (2D) \cite{Baxter}.   In 2D, the phase transition is of second order for $q\leq 4$ and of first order for higher $q$. In 3D, a number of points are well understood. For example, the phase transition is of first order for $q > 2$.  The molecular crystal described above undergoes therefore  a first-order transition in the absence of a dipolar interaction.

The purpose of this paper is to investigate the effect of the dipolar interaction on the ground state (GS) structure and on the nature of the phase transition. The dipolar interaction is a long-range interaction which yields different GS structures depending on the shape of the sample as will be discussed in the next section.  To carry out our purpose, we use a steepest descent method for the GS determination and the Monte Carlo (MC) simulation combined with the histogram technique to distinguish first- and second-order characters.

In section \ref{model}, we show our model and analyze the GS.  Results of MC simulations are shown  and discussed in section \ref{result}.
Concluding remarks are given in section \ref{conclu}.

\section{Model and Ground-State Analysis}\label{model}
We consider a simple cubic lattice where each site is occupied by an axial molecule.  The molecular axis can be in the $x$, $y$ or $z$ direction. Let us denote the orientation of the molecule at the lattice site $i$ by a unit segment, not a vector,  which can lie in the $x$, $y$ or $z$ direction.
We attribute the Potts variable $\sigma$=1, 2 or 3 when it lies in the $x$, $y$ or $z$ axes, respectively, in the calculation. By the very nature of the model shown below, the results do not depend on these numbers as in the standard $q$-state Potts model.

 We suppose that the interaction energy between two nearest molecules is -$J$ ($J>0$) if their axes are parallel, zero otherwise. With this hypothesis, the  Hamiltonian is given by the following 3-state Potts model:

\begin{equation}\label{HL}
{\cal H} = -J\sum_{(i,j)}\delta(\sigma_i,\sigma_j )
\end{equation}
where $\sigma_i$ is the 3-state Potts variable at the lattice site $i$ and $\sum_{(i,j)}$ is made
over the nearest sites $\sigma_i$ and $\sigma_j$.

The dipolar interaction between  Potts variables   is written as
\begin{equation}\label{dip}
{\cal H}_d=D\sum_{(i,j)}\{\frac{1}{r_{ij}^3}-
3\frac{[\mathbf{S}(\sigma_i)\cdot \mathbf u_{ij}][\mathbf{S}(\sigma_j)\cdot \mathbf u_{ij}]}{r_{ij}^3}\}\delta (\sigma_i,\sigma_j )
\end{equation}
where $\mathbf u_{ij}$ is the vector of unit length connecting sites $i$ and $j$, $D$ a positive constant depending on the material, $\mathbf{S}(\sigma_i)$ is defined as the unit vector lying on the axis corresponding to the value of $\sigma_i$.
The sum
is limited at some cutoff distance $r_c$.  Without the Kronecker condition and in the case of classical XY or Heisenberg spins, the dipolar interaction gives rise to spin configurations which depend on the sample shape. For example, in 2D or in rectangular slabs spins lie in the plane to minimize the system energy.

Let us first  discuss about the GS in the Potts model introduced above. When $D=0$ the GS is uniform with one orientation value, namely it is  3-fold degenerate.  However, for a  nonzero $D$, the ground state changes with varying $r_c$.
To determine the GS, we use the steepest descent method which works very well in systems with uniformly distributed interactions.  This method is very simple \cite{Ngo2007,Ngo2007a} (i) we generate an initial configuration at random (ii) we calculate the local field created at a site by its neighbors using  (\ref{HL}) and   (\ref{dip}) (iii) we take the Potts variable to minimize its energy (i. e. align the "Potts spin" in its local field) (iv) we go to another site and repeat until all sites are visited: we say we make one sweep (v) we  do a large number of sweeps per site until a good convergence is reached.

\begin{figure}[h!]
 \centering
 \includegraphics[width=50mm,angle=0]{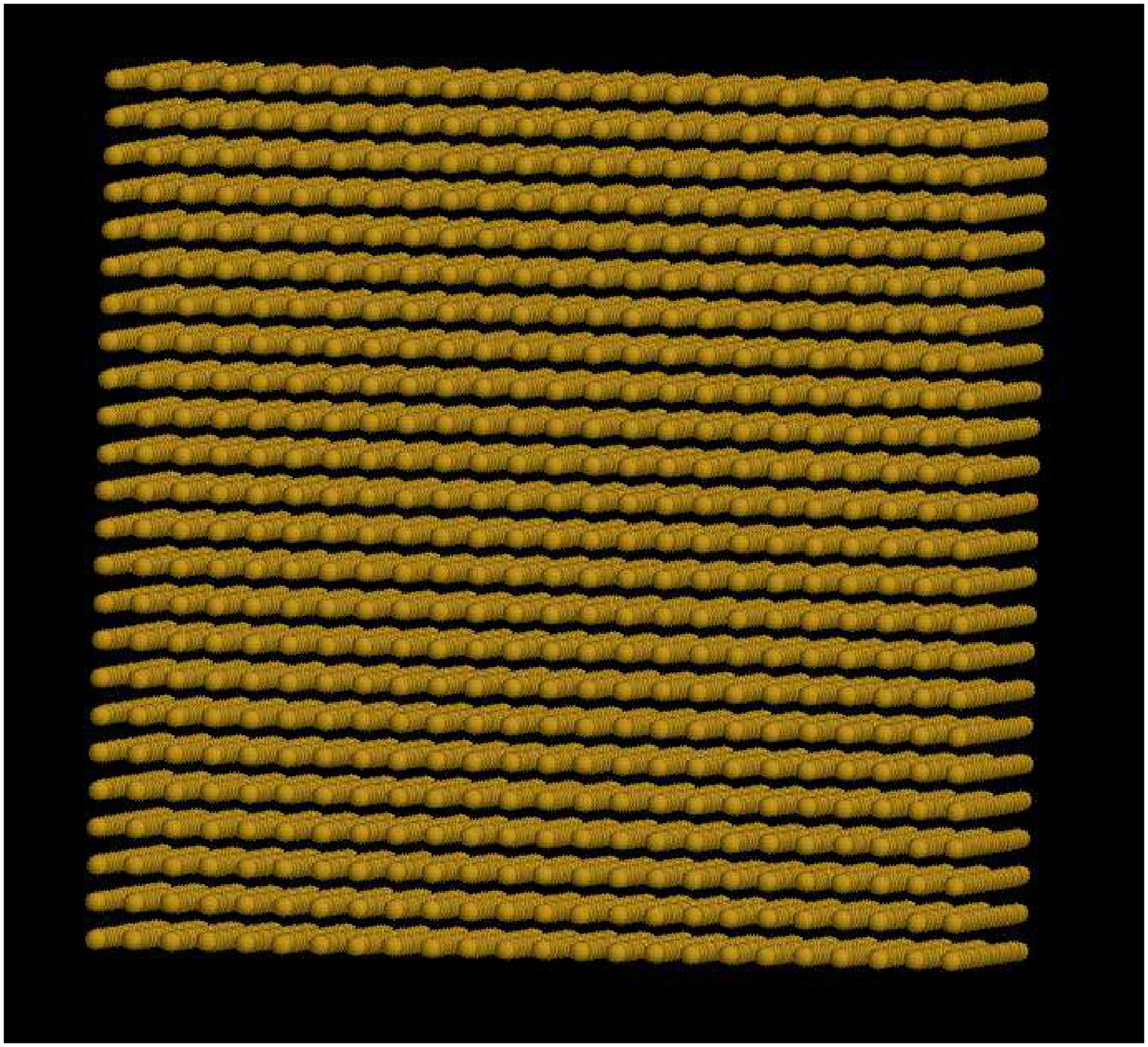}
 \includegraphics[width=50mm,angle=0]{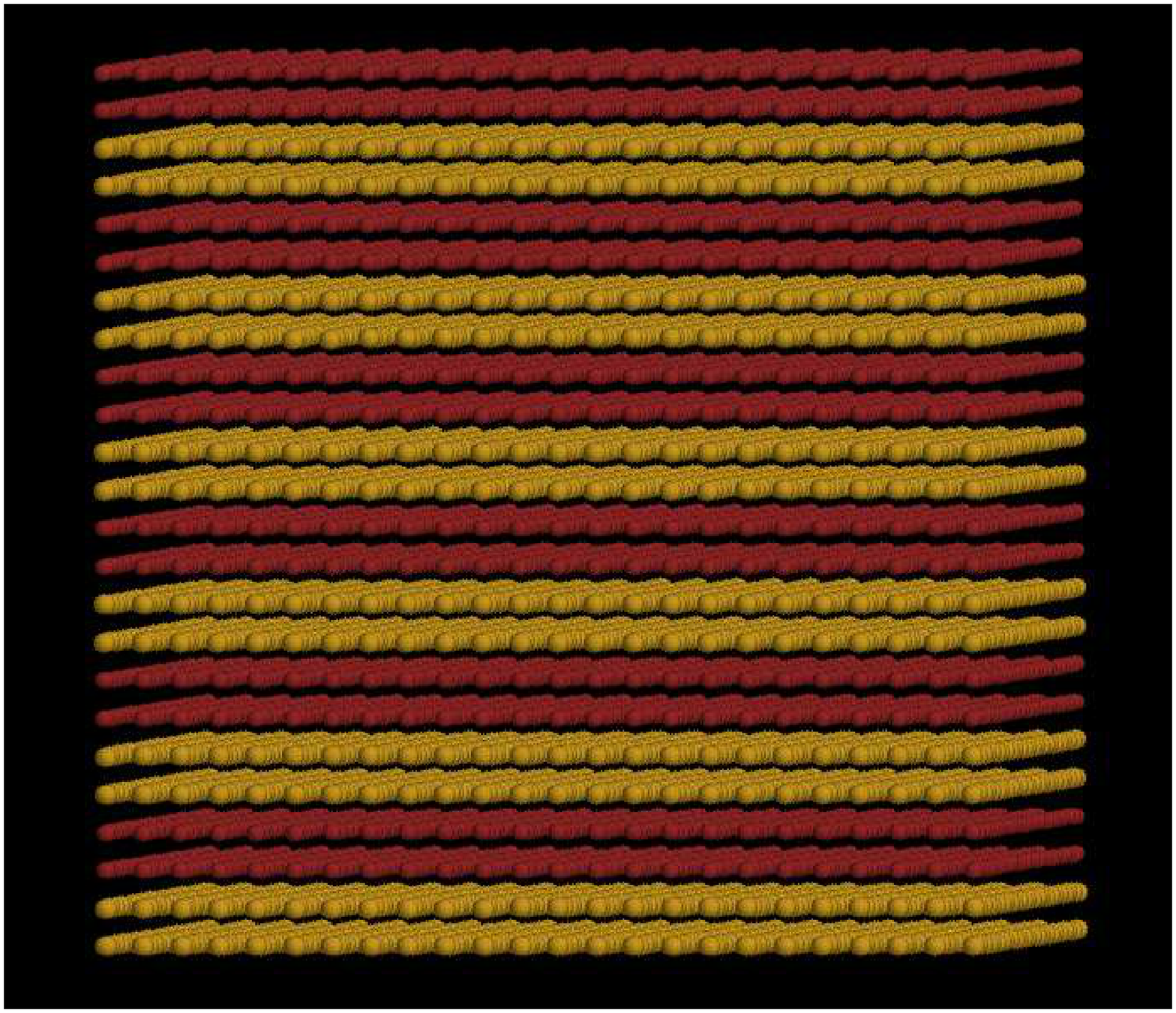}
 \includegraphics[width=50mm,angle=0]{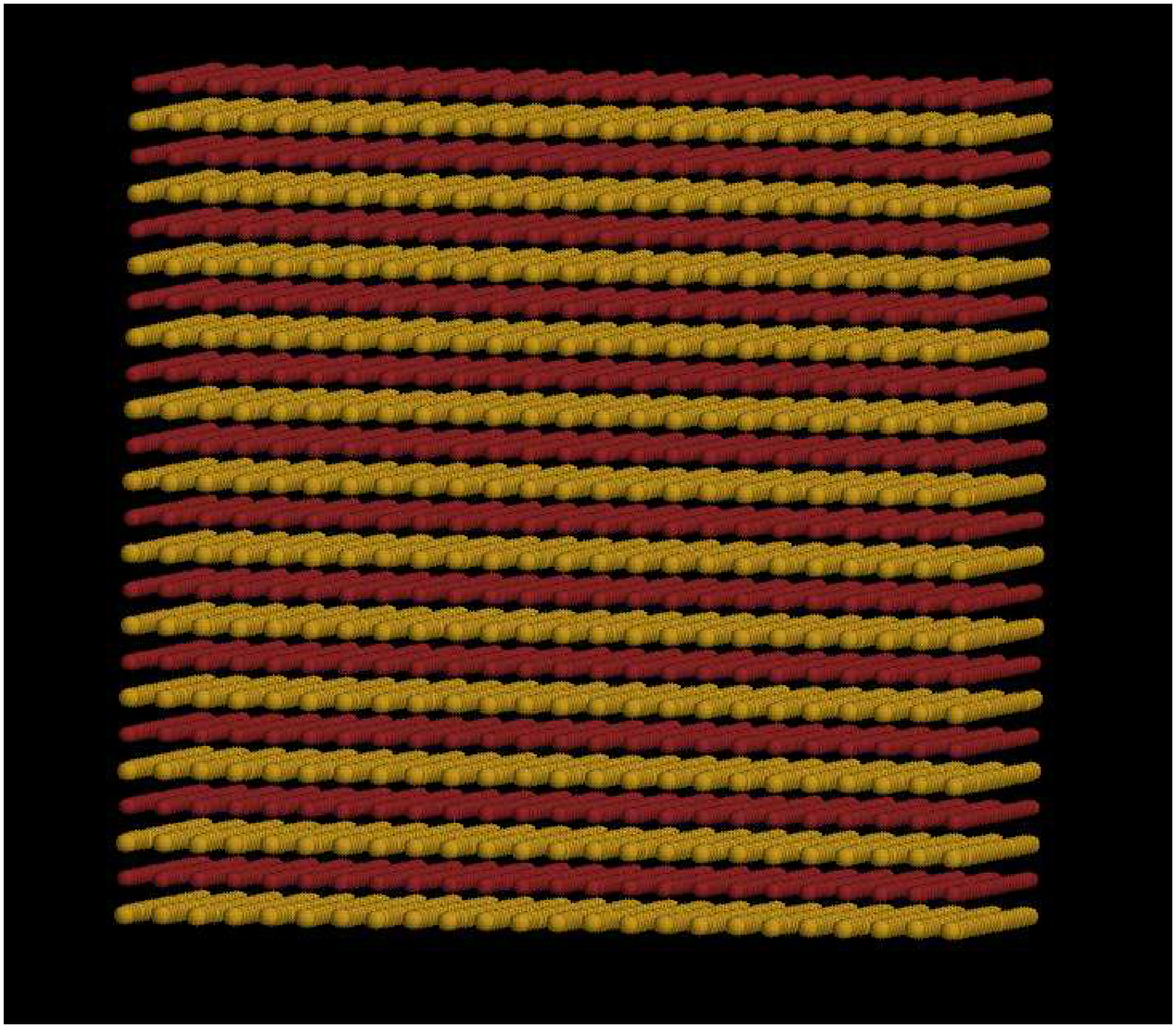}
 \includegraphics[width=50mm,angle=0]{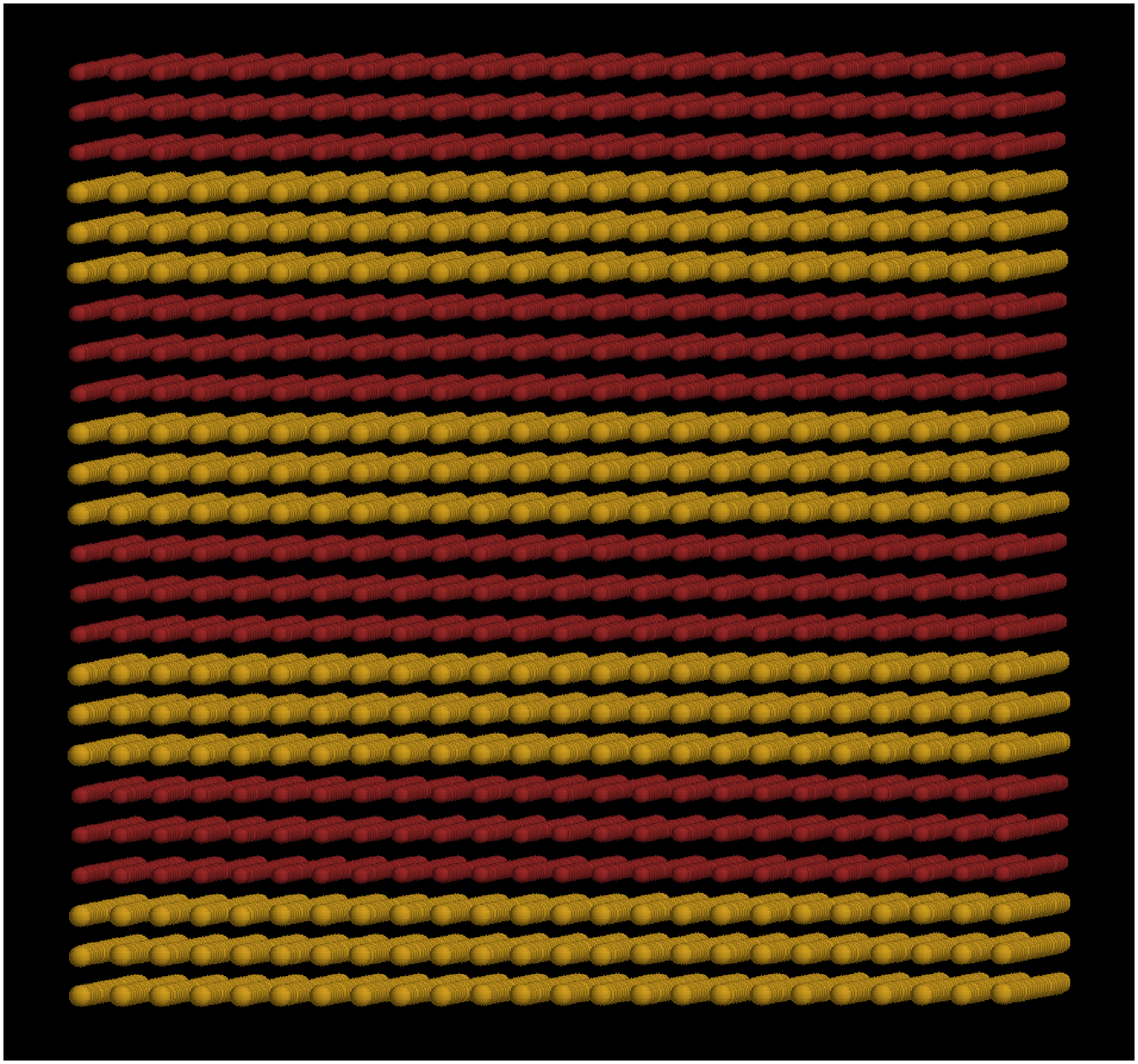}
 \caption{Ground state in the case where a) $D/J=0.4$, $r_c=2.3$: uniform configuration ; b) $D/J=1$, $r_c=2.3$: double-layer structure; c) $D/J=2$, $r_c=2.3$: single-layer structure; d) $D/J=0.4$, $r_c=\sqrt{10}\simeq 3.16$: triple-layer structure.  See text for comments.} \label{GS}
\end{figure}

We show some examples in Fig. \ref{GS}.  For very small $D$, the GS is uniform for any $r_c$ as shown in Fig. \ref{GS}a.   For increasing $D$, the GS has layered structures with period $p=1$ (alternate single layers), $p=2$ (double layers), $p=3$ (triple layers),... depending on $r_c$ and $D$.

We note that in Fig. \ref{GS} there are only two kinds of molecular orientations represented by two colors (on line) in spite of the fact that we have  three possible orientations.
Let us discuss about the single-layer structure shown in Fig. \ref{GS}c.  It is very important to note that each layer has no coupling with two nearest layers whose molecules lie on another axis. However, it is coupled to two next-nearest layers of the same color, namely the same molecular axis,  lying within $r_c$.
In other words, the GS is composed of two interpenetrating "independent" subsystems.  The same is true for other layered structures: in a double-layer structure molecules of the same orientation (same color on line) interact with each other but they are separated by a double layer of  molecules of another orientation.
To our knowledge, this kind of GS has never been found before.  It may have important applications
at macroscopic levels.

 Note also that if the molecules  of the  single-layer structure are successively in $x$-oriented  and $y$-oriented planes, then the stacking direction of these planes is the $z$ direction.  The molecules can choose the $x$ and $z$ orientations or the $y$ and $z$ directions. In those cases the stacking directions are respectively the $y$ and $x$ directions.   These three possibilities are equivalent if the sample is cubic.   In rectangular shapes, the stacking direction is along the smallest thickness.

In order to understand how such GS configurations found by the steepest descent method depend on $D$ and $r_c$, we have considered the structures 121212...(single-layer structure), 11221122... (2-layer structure), 111222111222... (3-layer structure)  and carried out the calculations of the energy of  a molecule $\sigma_i$ interacting with its neighbors $\sigma_j$.  The case where the configuration is uniform, i. e. there is only one kind of molecular orientation,  say  axis $x$, the dipolar energy of  $\sigma_i$ is
\begin{equation}\label{Ei}
E_i=D\sum_j [\frac{1}{r_{ij}^3}-3\frac{(u_{ij}^x)^2}{r_{ij}^3}]
\end{equation}
Note that with the use of the Potts model for the dipolar term in Eq. (\ref{dip}), the energy depends only on the axis, not on its direction as seen by the square term $(u_{ij}^x)^2$.  In this sense, the model is suitable to describe   axial, but non-directed, interacting molecules.

If we transform the sum in integral,  the sum in the first term gives $4\pi \ln r_c$ (integrating from 1 to $r_c$), while the second term gives -4$\pi \ln r_c$, which cancels the first term. This is valid  for $r_c$ larger than 1.
Thus the dipolar energy $E_d=D[4\pi\ln r_c-4\pi \ln r_c]$  is zero for the uniform configuration in 3D space in the framework of our model.  The energy of the system comes from the short-range exchange term, Eq. (\ref{HL}).

\begin{figure}[h!]
 \centering
 \includegraphics[width=50mm,angle=0]{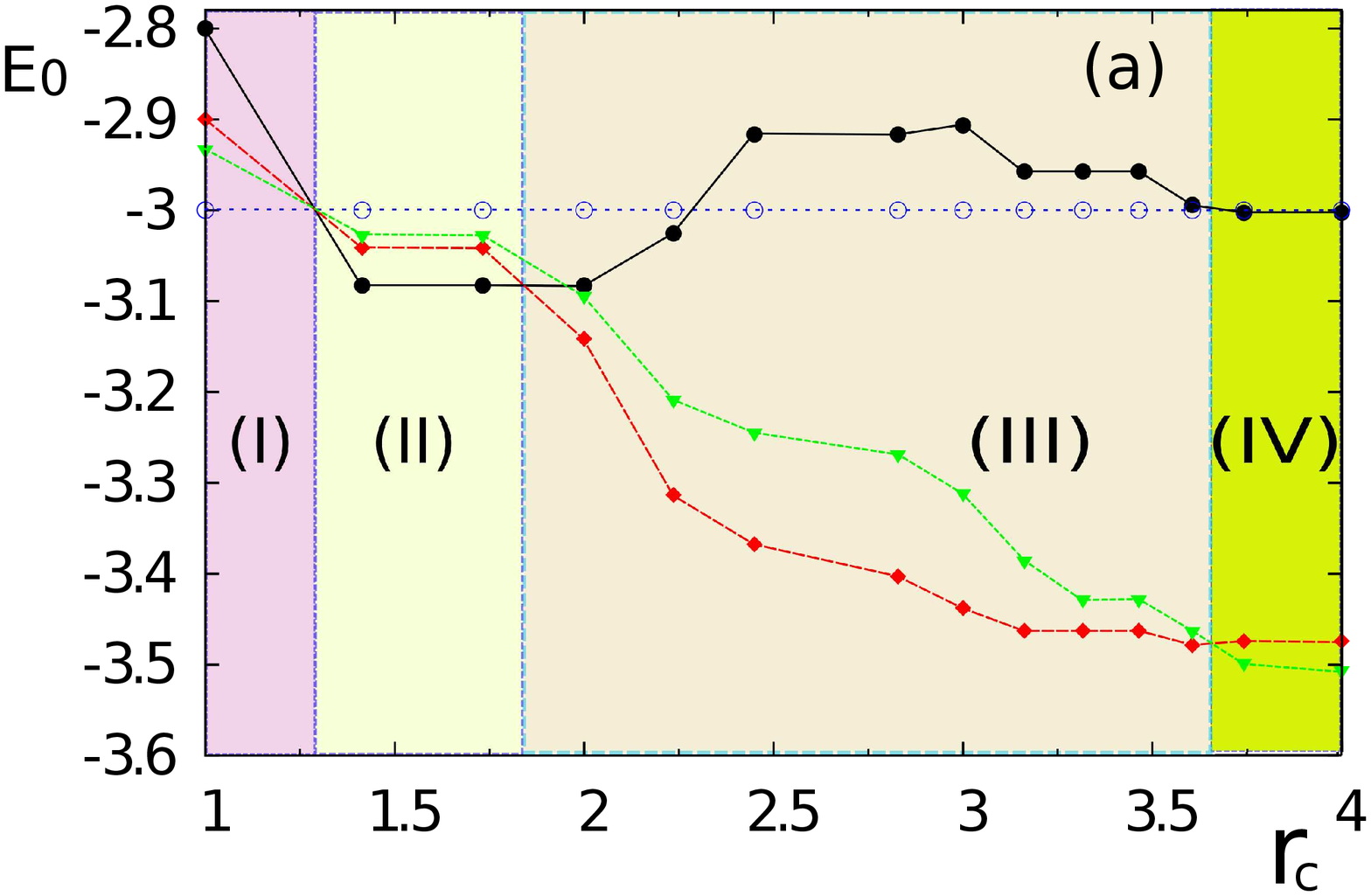}
 \includegraphics[width=50mm,angle=0]{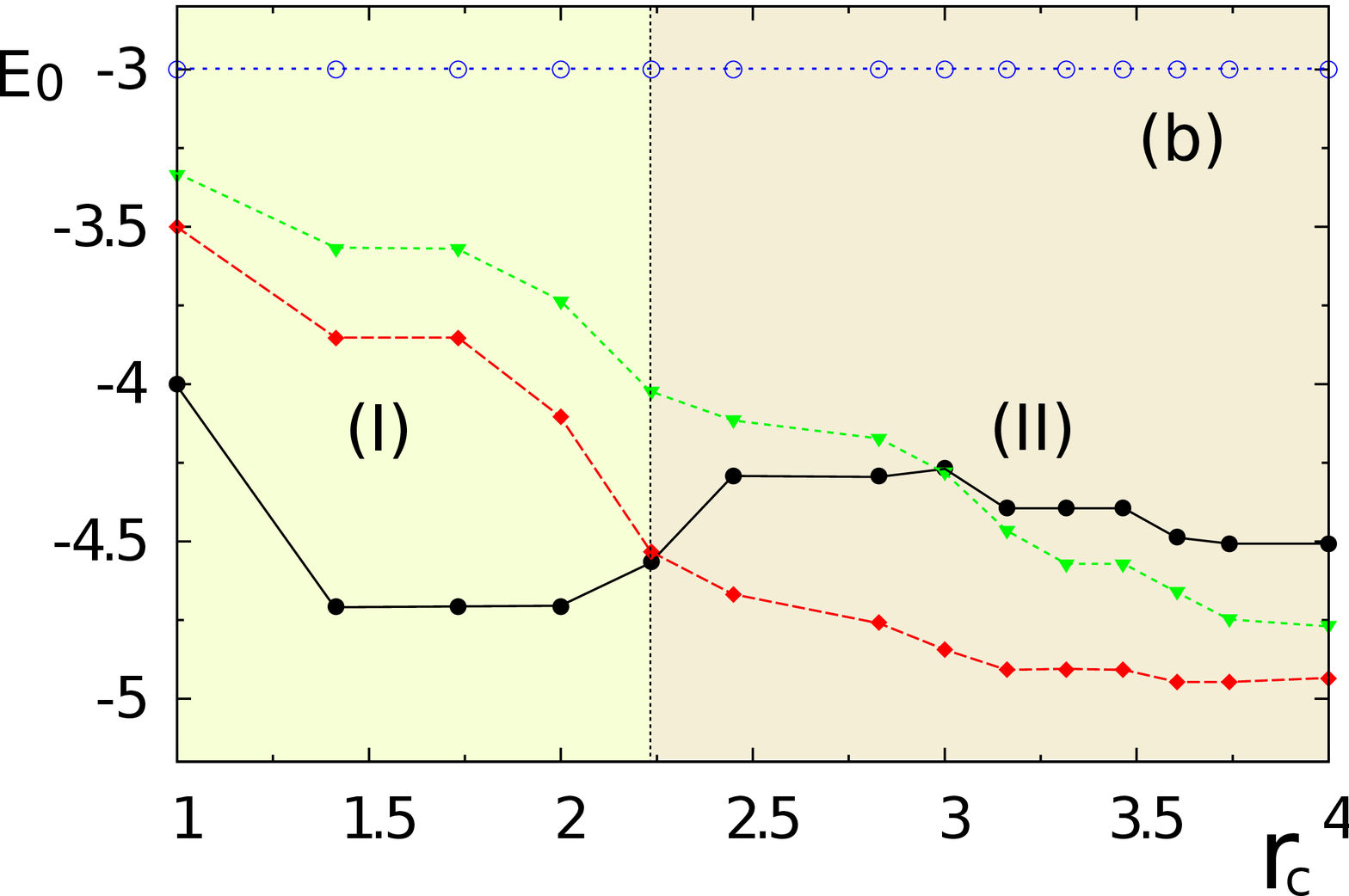}
 \caption{Ground state energy versus $r_c$ in the case where (a) $D/J=0.8$ (b) $D/J=2$. Blue, black, red and green lines  represent GS energy of the uniform,  single-layer,  double-layer and triple-layer structures, respectively. Zones I, II, II,  IV indicate these respective different GS.  See text for comments.} \label{GSER}
\end{figure}

The energy per site of other layered structures with periodicity $p=2$ and 3 is numerically calculated and shown in Fig. \ref{GSER} for $D/J=0.8$ and 2.   In each figure, the GS is the structure which corresponds to the lowest energy. One sees  in Fig. \ref{GSER}a the following GS configurations with varying $r_c$:

- uniform GS:  for $1 \leq r_c \leq 1.3$ (zone I)

- single-layer structure:  for $1.3 \leq r_c \leq 1.8$ (zone II)

- double-layer structure: for $1.8 \leq r_c \leq 3.65$ (zone III)

- triple-layer structure: for $ r_c \geq 3.65$ (zone IV)

For $D/J=2$ there are only two possible GS with varying $r_c$ as shown in Fig. \ref{GSER}b.

\begin{figure}[h!]
 \centering
 \includegraphics[width=50mm,angle=0]{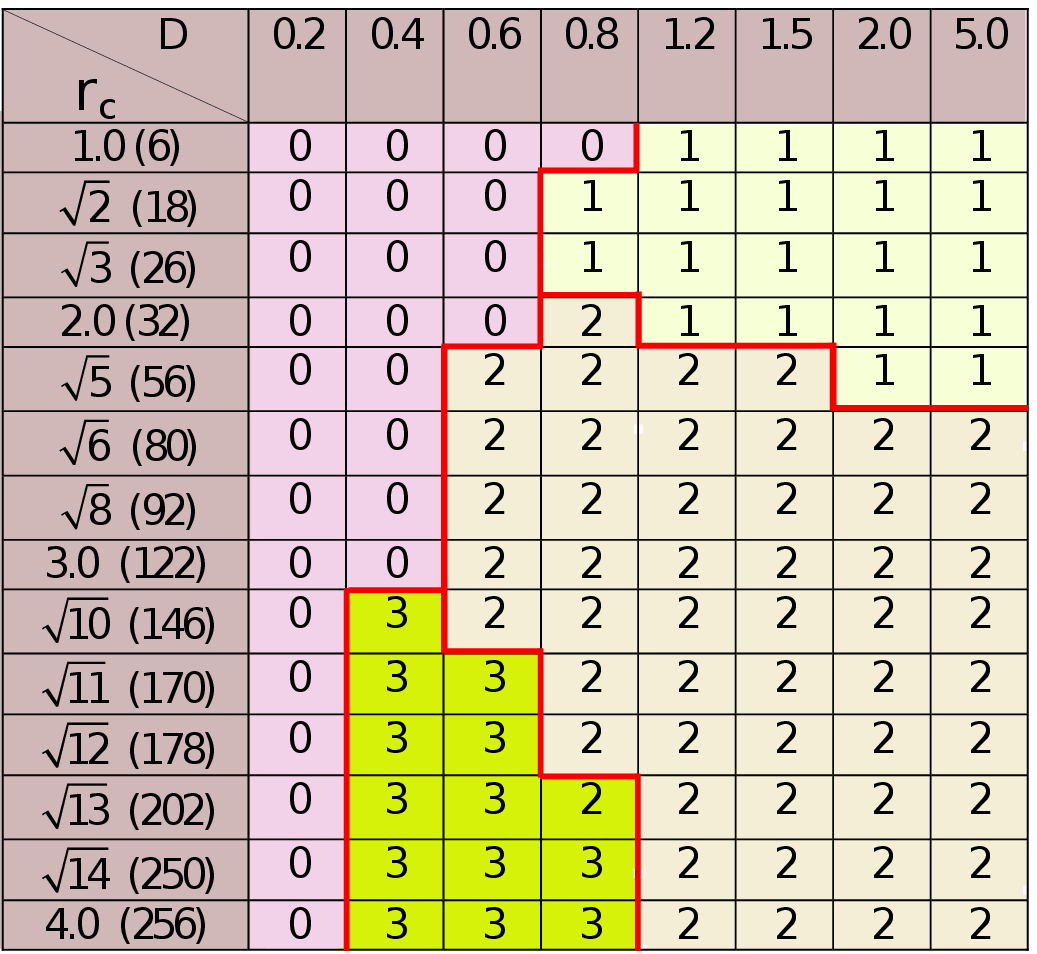}
 \caption{Ground state in space ($r_c$,$D$): the numbers 0, 1, 2 and 3 denote the uniform, single-layer, double-layer and triple-layer structures, respectively.  The first column displays the values of $r_c$ with the number of neighbors indicated in the parentheses.} \label{GSDR}
\end{figure}

We summarize in Fig. \ref{GSDR} the different GS in the space ($r_c,D$) with $J=1$. We note that for a given $D$, for example $D=0.8$, the GS configuration starts with uniform configuration then with period 1, 2, 3, ... for increasing $r_c$.  At large $D$, uniform configuration is not possible at any $r_c$.  Long-period configurations are on the other hand favored at small $D$ and large $r_c$.

\section{Phase Transition: Results}\label{result}

We consider  a sample size of $N\times N\times N_z$ where $N$ and $N_z$ vary from 24 to 48 but $N_z$ can be different from $N$ in order to detect the shape-dependence of the ground state..  The 3-state Potts model with NN exchange interaction $J=1$ is used to describe the three molecular orientations.  For the dipolar term, a cutoff distance $r_c$ is taken  up to $\sqrt{10}\simeq 3.16$ lattice distance. At this value, each molecule has a dipolar interaction with  146 neighbors.   Periodic boundary conditions in all directions are employed.

We have used the standard MC method \cite{Binder} with the system size from $24^3$ to $48^3$.  The equilibrating time $N_1$ is about $10^6$ MC steps per site, and the averaging time $N_2$ is between $10^6$ and $10^7$ MC steps per site.  The averages of the internal energy $<U>$ and the specific heat $C_V$ are defined by

\begin{eqnarray}
 \langle U\rangle&=&<{\cal H} + {\cal H}_d>\\
C_V&=&\frac{\langle U^2\rangle-\langle U\rangle^2}{k_BT^2}
\end{eqnarray}
where $<...>$ indicates the thermal average taken over $N_2$ microscopic states at $T$.

We define the Potts order parameter $Q$ by
\begin{equation}\label{Q}
Q=[q\max (Q_1,Q_2,...,Q_q)-1]/(q-1)
\end{equation}
where $Q_n$ is the spatial average defined by
\begin{equation}\label{Qn}
Q_n=\sum_j \delta (\sigma_j-n)/(N\times  N\times N_z)
\end{equation}
$n(n=1,...,q)$ being the value of the Potts variable at the site $j$.  For $q=3$, one has $n=1,2,3$ representing respectively the molecular axis in the $x$, $y$ and $z$ directions.
The susceptibility is defined by
\begin{equation}\label{chi}
\chi=\frac{\langle Q^2\rangle-\langle Q\rangle^2}{k_BT}
\end{equation}

In MC simulations, we work at finite sizes, so for each size we have to determine the "pseudo" transition which corresponds in general to the maximum of the specific heat or of the susceptibility. The
maxima of these quantities need not to be at the same temperature. Only at the infinite size, they should coincide. The theory of finite-size scaling \cite{Hohenberg,Ferrenberg1,Ferrenberg2} permits to deduce properties of a system at its thermodynamic limit.  We have used in this work a size large enough to be close to the extrapolated bulk transition temperature.

\begin{figure}[h!]
 \centering
 \includegraphics[width=60mm,angle=0]{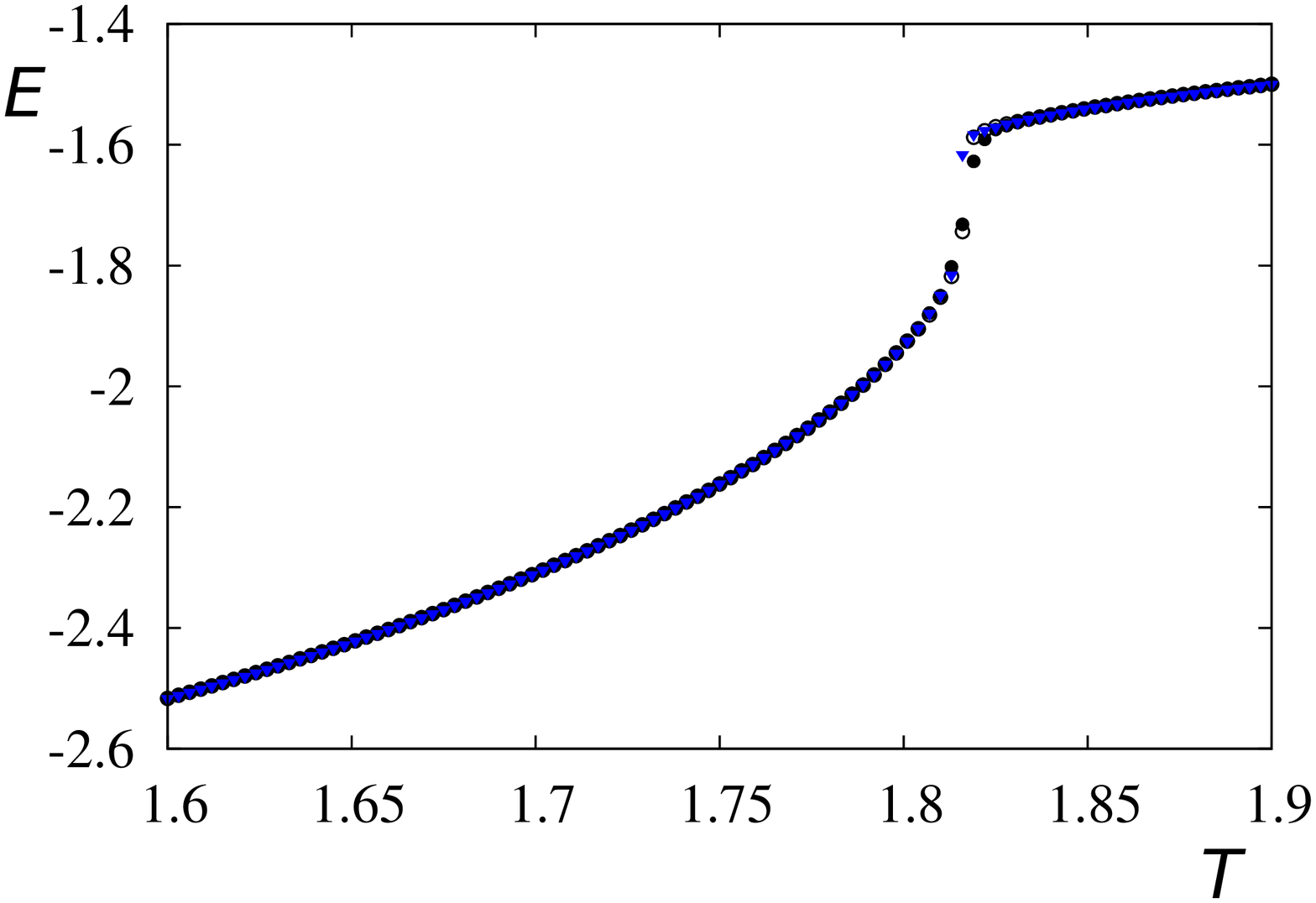}
  \includegraphics[width=60mm,angle=0]{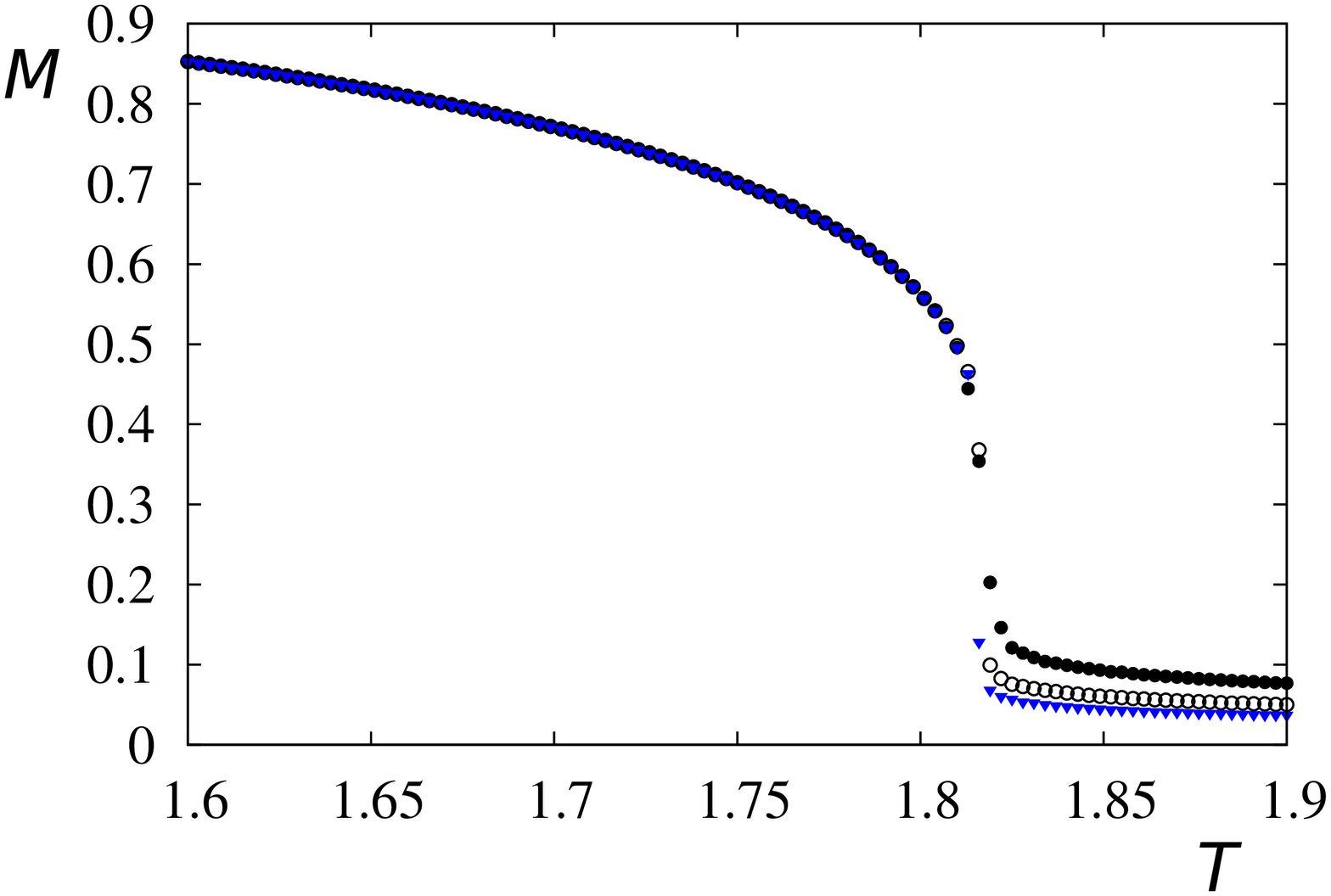}
 \caption{Internal energy  per spin $E$ (upper) and order parameter $M=<Q>$  (lower) versus temperature $T$ without dipolar interaction, for several sizes $N_z=N=24$ (black circles), $36$ (void circles), $48$ (blue triangles, color on line), $J=1$.} \label{ET}
\end{figure}

\begin{figure}[h!]
 \centering
 \includegraphics[width=60mm,angle=0]{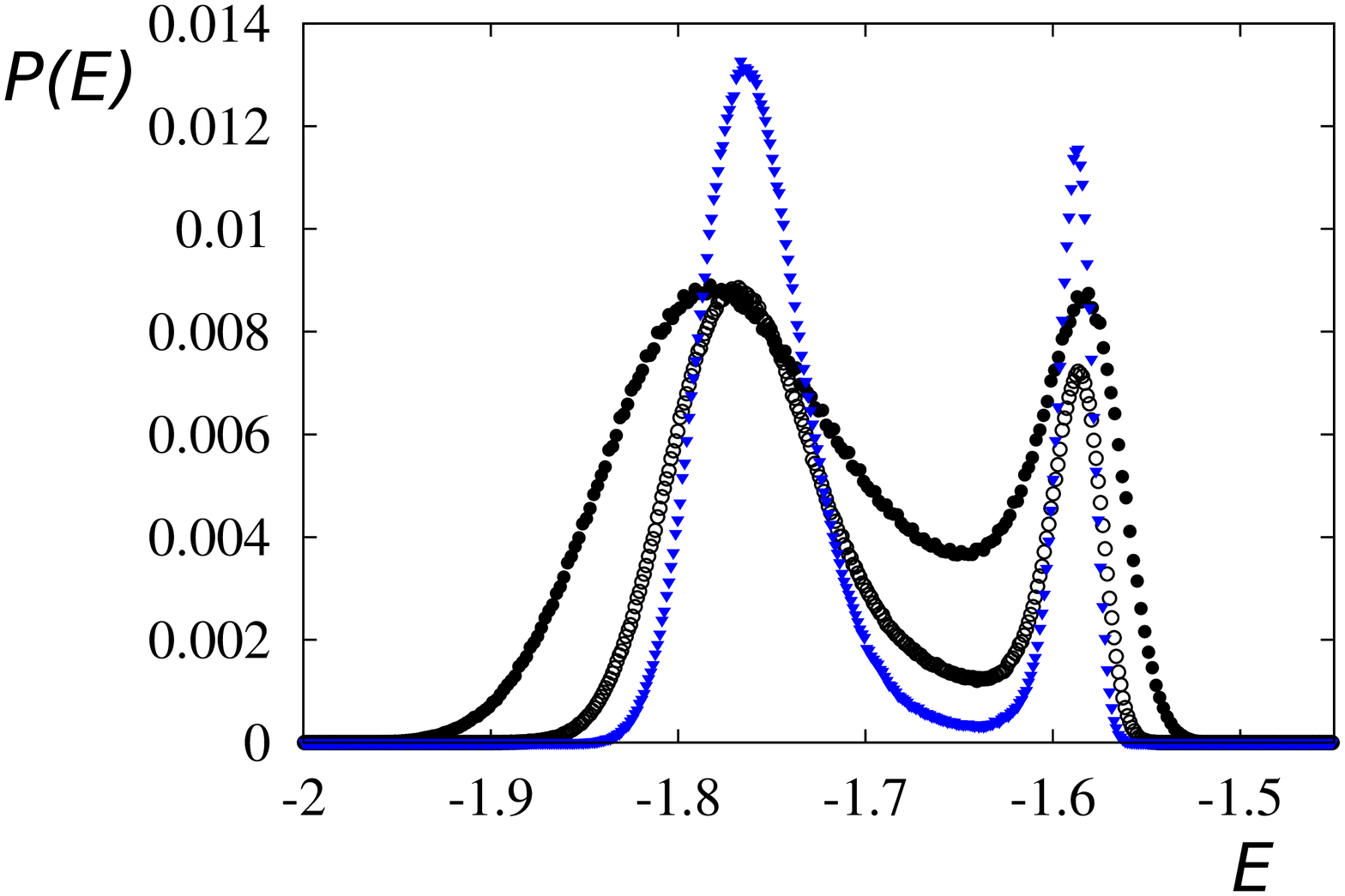}
  \includegraphics[width=60mm,angle=0]{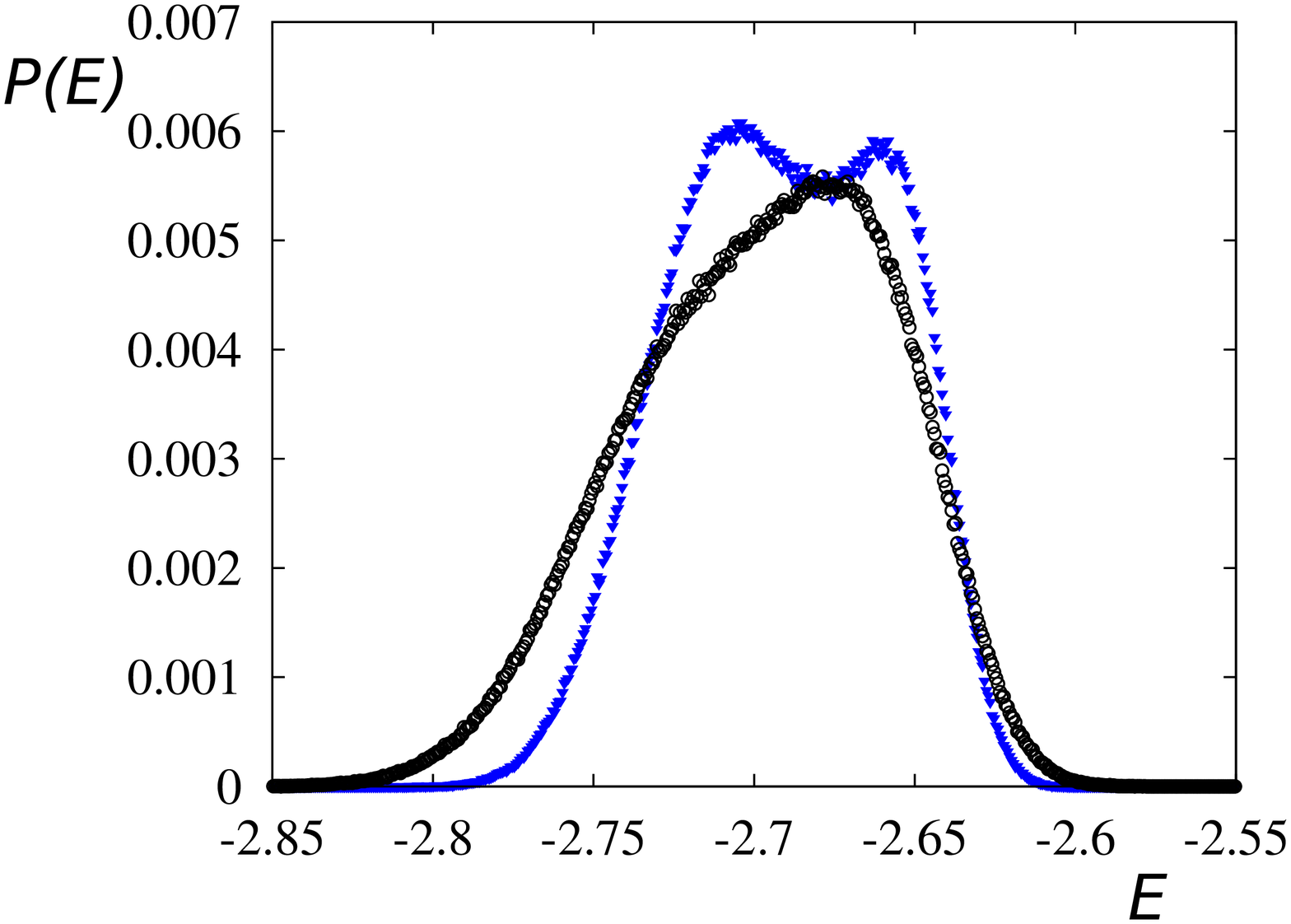}
 \caption{Energy histogram at the transition temperature without ($D=0$, upper) and with ($D=2$, lower) dipolar interaction, for several sizes $N_z=N=24$ (black circles), $36$ (void circles), $48$ (blue triangles, color on line), $r_c=\sqrt{10}$,  $J=1$.} \label{PD0}
\end{figure}

We show first in Fig. \ref{ET} the energy per site $E\equiv <U>/(N\times N\times N_z)$ and
the order parameter $M=<Q>$ versus $T$,  for several lattice sizes in the absence of the dipolar interaction, i. e. $D=0$.  As said earlier this case corresponds to the 3-state Potts model.  We find a very sharp transition. In order to check the first-order nature of this transition, we used the histogram technique which is very efficient in detecting weak first-order transitions and in calculating critical exponents of second-order transitions \cite{Ferrenberg1,Ferrenberg2}.  The main idea of this technique is to make an energy histogram at a temperature $T_0$ as close as possible to the transition temperature. Often, one has to try at several temperatures in the transition region.  Using this histogram in the formulae of statistical physics for canonical distribution, one obtains energy histograms in a range of temperature around $T_0$.   In second-order transitions, these histograms are gaussian. They allows us to calculate averages of physical quantities as well as critical exponents using the finite-size scaling.  In first-order transitions, the energy histogram shows a double-peak structure at large enough lattice sizes.
Using this method, we have calculated the histogram shown in Fig. \ref{PD0} for several lattice sizes. With increasing size, the two peaks are well separated with the dip going down to zero, indicating  a tendency toward an energy discontinuity at the infinite size.  The distance between the two peaks is the latent heat $\Delta E$.

We take into account now the dipolar interaction. In order to see its progressive effect, we  perform simulations with $r_c$ varying between $r_c=\sqrt{6}$ and $r_c=\sqrt{10}$ and follow
the change of the characteristics of the phase transition.

\begin{figure}[h!]
 \centering
   \includegraphics[width=60mm,angle=0]{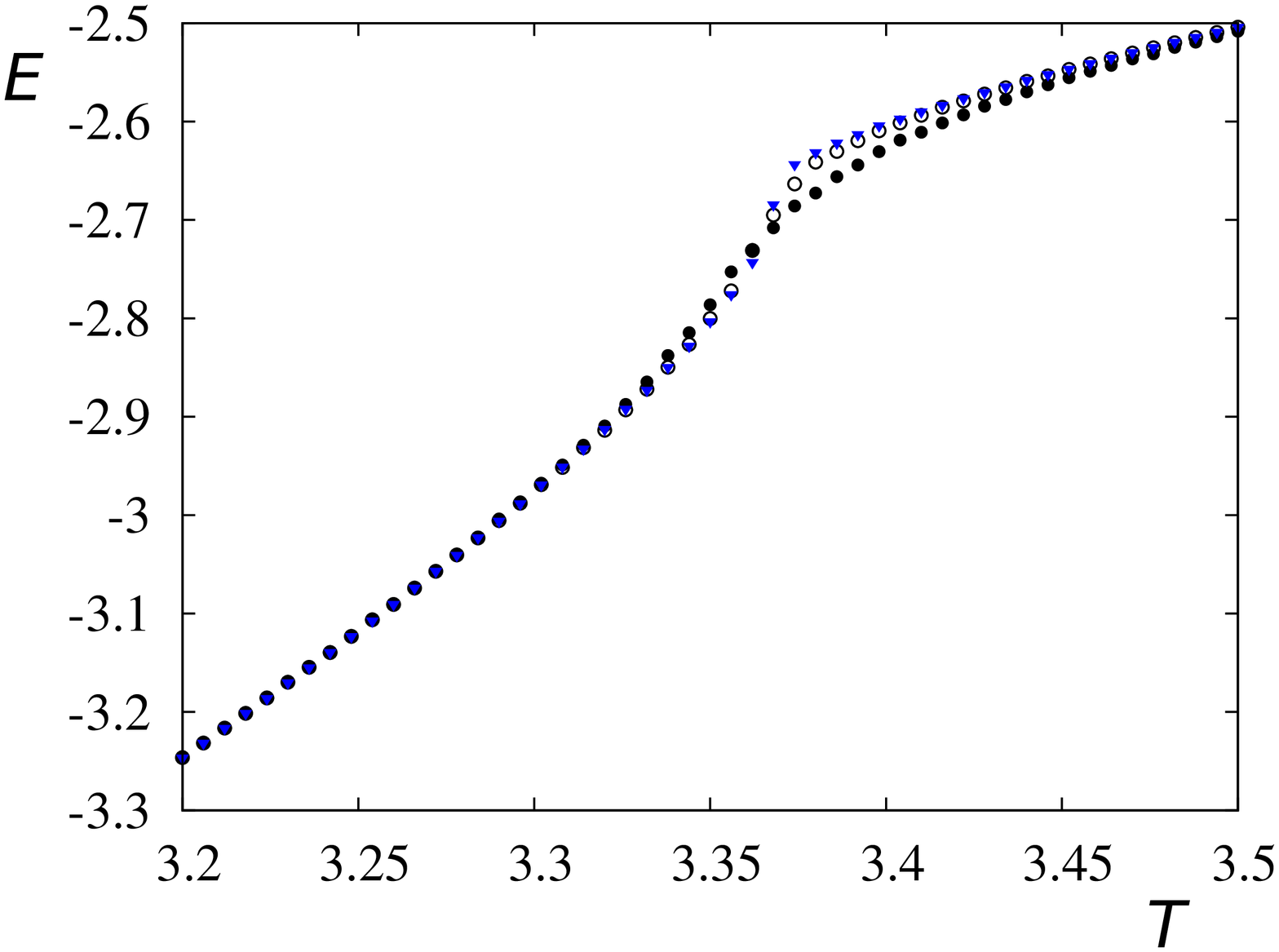}
   \includegraphics[width=60mm,angle=0]{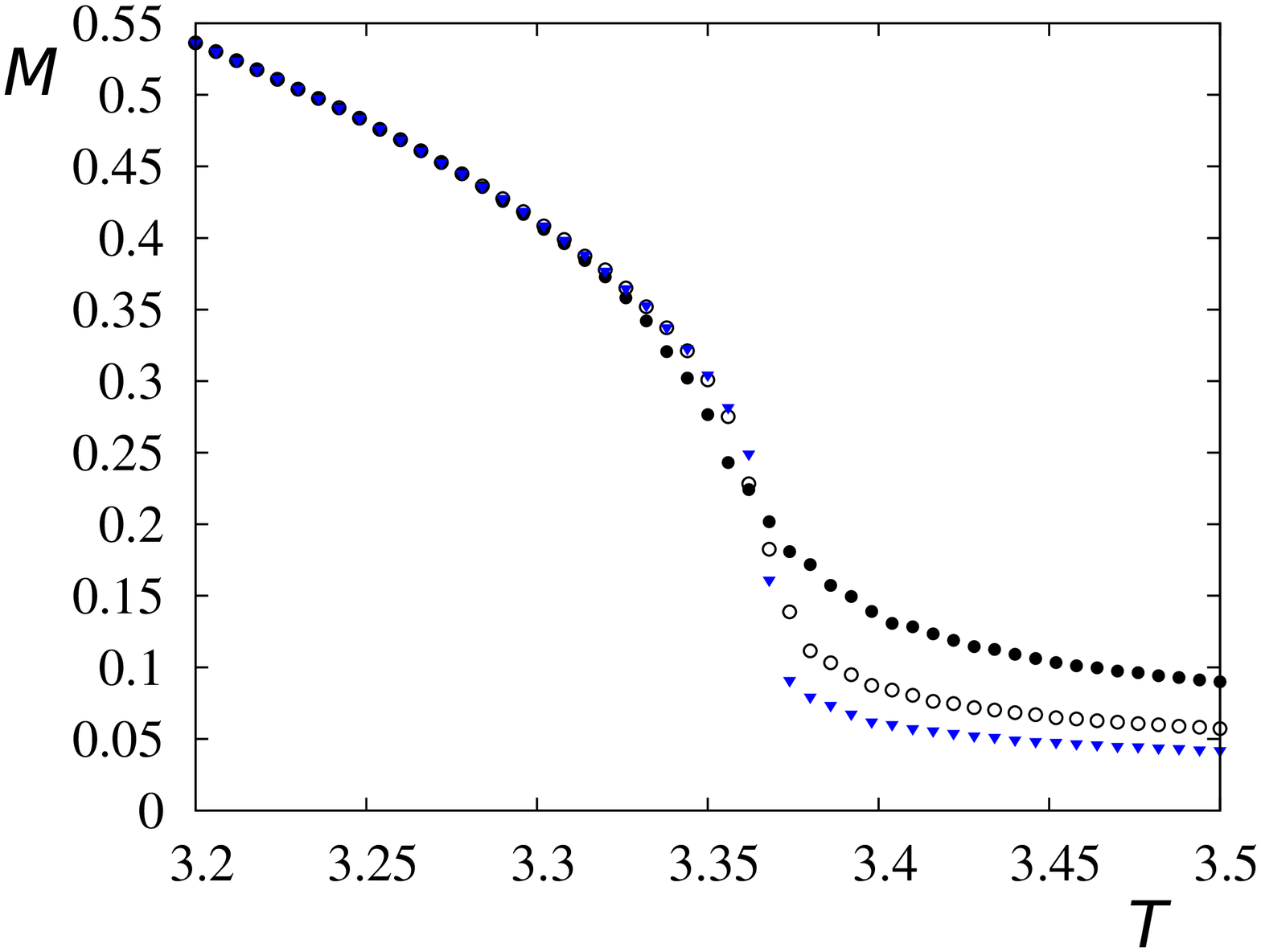}
 \caption{Internal energy per spin $E$ and order parameter $M=<Q>$ versus $T$ for $D=2$, $r_c=\sqrt{10}$ with several lattice sizes  $N_z=N=24$ (black circles), $36$ (void circles), $48$ (blue triangles, color on line), and $J=1$.} \label{EMD}
\end{figure}

We show in Fig. \ref{EMD} the energy and the order parameter versus $T$ for $D=2$, $r_c=\sqrt{10}$.
The transition is still very sharp but the strong first-order character diminishes.

\begin{figure}[h!]
 \centering
 \includegraphics[width=60mm,angle=0]{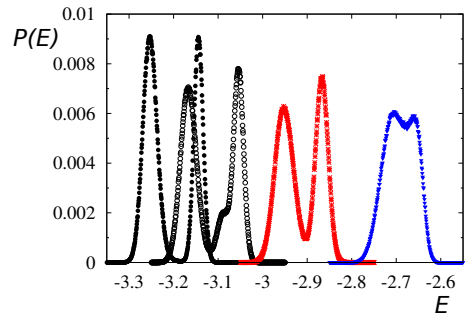}
 \caption{Energy histogram showing double-peak structure for $D=2$ at several values of $r_c$: from left to right $r_c=\sqrt{6}$ (black circles), $\sqrt{8}$ (void circles), 3 (red stars, color on line) and $\sqrt{10}$ (blue triangles, color on line). $N_z=N= 48$, $J=1$.} \label{PED}
\end{figure}

We show in Fig. \ref{PED} the energy histogram for several values of $r_c$  with $D=2$.  As seen the latent heat becomes small at $r_c=\sqrt{10}$.

We observe that the latent heat $\Delta E$ diminishes with increasing $r_c$. However, we cannot conclude that the first-order disappears at large $r_c$.  To check that point we need to go to larger $r_c$ which will take a huge CPU time because of the increasing number of neighbors (we recall that for $r_c=\sqrt{10}$, we have 146 neighbors for each molecules).
We observe from Fig. \ref{PED} that the latent heat does not change significantly with $r_c$ up to $r_c\simeq 2.85$, meaning that the first order transition is dominated by the short-range 3-state Potts interaction.  The latent heat decreases rather strongly afterward but we do not know if it tends to zero or not.




 \section{Conclusion}\label{conclu}

 We have studied in this paper a molecular crystal characterized by three possible orientations of the molecular axes.  The model is described by a short-range 3-state Potts model and a dipolar Potts interaction.  We have analyzed the ground state as functions of the dipolar interaction strength $D$ and its cutoff distance $r_c$.  The ground state shows  remarkable properties: it is composed of two interpenetrating independent subsystems with a layered structure whose periodicity depends on $D$ and $r_c$.  To our knowledge, such a ground state has not been seen in uniformly interacting molecules. The stacking of independent ordered layers is reminiscent of a smectic ordering. Note that the model is applied to smectic phases where tilted angles are uniform: it can be of type $A$ as supposed here or of type $C$ if the molecular axis does not coincide with the $x$, $y$ or $z$ axis of the lattice. The coupling between adjacent different ordered planes is here strictly zero at $T=0$, due to the Potts condition in the model.
 We have used the Monte Carlo histogram method to study the phase transition in this system.
 In the absence of the dipolar interaction, the model which is  a 3-state Potts model is known to undergo a first-order phase transition from the orientational ordered phase to the disordered state.   Upon the introduction of a dipolar interaction of sufficient strength into the Potts model, the ground state is broken into layers as described above.   We have shown that
 the transition remains  of first order as the cutoff distance is increased at least  up to $r_c=\sqrt{10}$.

{}

\end{document}